\documentclass[aps, prd, twocolumn, superscriptaddress, nofootinbib, floatfix, noshowpacs]{revtex4-2}
\usepackage{amsmath,placeins,graphicx,amssymb,tabularx}
\allowdisplaybreaks  
\usepackage[percent]{overpic}
\usepackage{xcolor}

\definecolor{purple}{rgb}{0.5,0,0.5}
\definecolor{blue}{rgb}{0.0,0,0.9}
\definecolor{prdblue}{rgb}{0.133,0.118,0.498}
\usepackage[colorlinks=true, pdfstartview=FitV, linkcolor=prdblue, citecolor= prdblue, urlcolor=prdblue]{hyperref}

\usepackage{orcidlink} 
\usepackage[markup=underlined]{changes}
\definechangesauthor[name={TK},color=blue]{TK}
\definechangesauthor[name={xpli},color=red]{xpli}

\begin{document}


\title{Crossover or First-Order: Impacts of Regularization}



\newcommand{\orcidauthorA}{\orcidlink{0009-0008-1946-1997}} 
\newcommand{\orcidauthorB}{\orcidlink{0009-0005-3091-393X}}
\newcommand{\orcidauthorC}{\orcidlink{0000-0003-3890-0242}}
\newcommand{\orcidauthorD}{\orcidlink{0000-0001-5445-9528}}

\author{Xin-Peng Li \orcidauthorA{}}
\email[]{xpli@smail.nju.edu.cn}
\affiliation{School of Physics, \href{https://ror.org/01rxvg760}{Nanjing University}, Nanjing, Jiangsu 210093, China}
\affiliation{Institute for Nonperturbative Physics, \href{https://ror.org/01rxvg760}{Nanjing University}, Nanjing, Jiangsu 210093, China}

\author{Hao-Ran Zhang \orcidauthorB{}}
\affiliation{School of Physics, \href{https://ror.org/01rxvg760}{Nanjing University}, Nanjing, Jiangsu 210093, China}
\affiliation{Institute for Nonperturbative Physics, \href{https://ror.org/01rxvg760}{Nanjing University}, Nanjing, Jiangsu 210093, China}

\author{Zhu-Fang Cui \orcidauthorC{}}
\email[]{phycui@nju.edu.cn}
\affiliation{School of Physics, \href{https://ror.org/01rxvg760}{Nanjing University}, Nanjing, Jiangsu 210093, China}
\affiliation{Institute for Nonperturbative Physics, \href{https://ror.org/01rxvg760}{Nanjing University}, Nanjing, Jiangsu 210093, China}

\author{Thomas Kl\"{a}hn \orcidauthorD{}}
\email[]{thomas.klaehn@csulb.edu}
\affiliation{Department of Physics $\&$ Astronomy,
\href{https://ror.org/0080fxk18}{California State University Long Beach}, Long Beach, CA 90840, USA}


\date{\today}

\begin{abstract}
The equation of state of hot, dense nuclear matter plays a fundamental role in many areas. 
However, owing to the nonperturbative nature of strong interactions, a reliable treatment is still under debate. 
We use a symmetry-preserving treatment of a vector\,$\otimes$\,vector contact interaction to study related issues at nonzero temperature or quark chemical potential, and carefully compare a noncovariant and a covariant regularization scheme. 
The numerical results show that the character of the phase transition depends sensitively on the regularization scheme and parameter choices.
\end{abstract}

\maketitle



\section{\label{Introduction}Introduction}
The Standard Model of particle physics stands as the most successful theoretical framework in modern high-energy physics, having been subjected to and passed a wide range of experimental tests over the past several decades. It provides a unified description of three of the four known fundamental interactions in nature:  electromagnetic, weak, and strong interactions. Among these, the strong interaction, which is described by the theory of quantum chromodynamics (QCD), is of particular importance. QCD is a non-Abelian gauge theory based on the $SU(3)_c$ color group, where quarks interact via the exchange of gluons. While perturbation theory has been proven to be extremely powerful at high energies, where the coupling constant becomes weak owing to asymptotic freedom, the situation changes dramatically at low energies. In the infrared regime, the QCD coupling grows large~\cite{Binosi:2016nme,Cui:2019dwv,dEnterria:2022hzv,Deur:2022msf,Deur:2023dzc} and renders perturbative methods ineffective, thereby necessitating alternative nonperturbative approaches.

The thermodynamics of strongly interacting matter, encapsulated in the equation of state (EoS) of QCD, plays a fundamental role in both cosmology and astrophysics. In cosmology, the QCD EoS governs the dynamics of the early universe during the quark-hadron transition, possibly influencing the evolution of primordial fluctuations and thus potentially leaving observable imprints in the cosmic microwave background or relic abundances~\cite{witten1984cosmic,applegate1985relics,schmid1997peaks}. In astrophysics, it dictates the internal structure and stability of compact stellar objects, such as neutron stars and hybrid stars. Accurate knowledge of the EoS at high baryon density, or chemical potential ($\mu$), is essential for determining properties such as maximum stellar masses, radii, and tidal deformabilities, which can now be constrained by multimessenger observations, including gravitational-wave detections from binary neutron star mergers~\cite{annala2018gravitational,raithel2019constraints}. Thus, a quantitative and reliable determination of the QCD phase structure and its corresponding EoS is a central challenge of modern theoretical physics.

Despite extensive progress achieved through lattice QCD simulations, serious limitations remain. At vanishing chemical potential, lattice QCD has provided quantitatively reliable results for the EoS, the pseudo-critical temperature of the crossover transition, and fluctuations of conserved charges. However, at nonzero $\mu$, lattice methods are plagued by the sign problem that arises from the complex phase of the fermion determinant~\citep{philipsen2013qcd}. This makes straightforward Monte Carlo sampling impractical and severely restricts applicability in the low-temperature $T$, high-$\mu$ domain, precisely the regime of relevance for neutron star interiors and heavy-ion collisions at lower beam energies. While methods such as reweighting, Taylor expansions, or analytic continuation from imaginary chemical potential have been explored, none of them yield a fully controlled and systematically improvable approach at large $\mu$~\cite{de2010simulating,nagata2022finite}.

In light of these obstacles, a variety of effective models have been proposed to capture essential features of QCD in regimes inaccessible to lattice simulations. Among them, for example, the Nambu-Jona-Lasinio (NJL) model and its Polyakov-loop extended version (PNJL) have been widely employed to investigate chiral symmetry breaking, its restoration at nonzero $T$ and $\mu$, as well as qualitative aspects of the phase diagram. The NJL model successfully incorporates spontaneous chiral symmetry breaking and dynamical mass generation, which are hallmarks of low-energy QCD~\cite{klevansky1992nambu,Klahn:2013kga,Xin:2014dia,Cui:2014hya,lu2015critical,Shi:2015ufa,Wang:2016fzr,Aoki:2017rjl,Shao:2017yzv,Khunjua:2018jmn,Cui:2018bor,Abreu:2019czp,Cui:2022vsr,Pasqualotto:2023hho,Zhang:2024dhs,Gifari:2024ssz}. Nevertheless, it also has well-known shortcomings: the absence of confinement, dependence on ultraviolet cutoff schemes that violate Poincar\'e covariance, and a strong sensitivity of predictions to parameter choices and regularization prescriptions~\cite{inagaki2015regularization,kohyama2016parameter}. As a result, while the NJL framework provides valuable intuition, its quantitative reliability remains questionable.

On the other hand, a popular first-principles continuum, nonperturbative approach that preserves fundamental symmetries is the Dyson-Schwinger equations (DSEs), which offer a system of coupled integral equations for Green's functions that are valid across all energy scales~\cite{roberts1994dyson,fischer2006infrared,roberts2000dyson,Fischer:2018sdj}. In particular, DSEs respect Poincar\'e covariance and allow for the simultaneous study of confinement, dynamical chiral symmetry breaking, and their interrelation at zero $T$ and $\mu$. While truncations of the infinite tower of equations are unavoidable, systematic schemes such as the rainbow-ladder approximation and beyond have been proven capable of producing results consistent with both lattice QCD and experimental observations in many contexts; for example, Refs.~\cite{wang2013baryon,Zhao:2014oha,Wang:2015tia,Xu:2015vna,Klahn:2015mfa,Cui:2016zqp,Gao:2016qkh,Li:2017zny,Isserstedt:2019pgx,Raya:2019dnh,Klaehn:2021nyv}.

In this work, we adopt a rainbow-truncated quark DSE with a local contact kernel to explore features of the QCD phase structure at nonzero $(T,\mu)$. 
In this set of approximations, the bulk thermodynamics is practically mean-field equivalent to NJL-type models \cite{Klahn:2015mfa}.
Crucially, because the contact interaction is nonrenormalizable, the regulator is part of the model. As we demonstrate, this leads to situations where the choice of the regularization scheme---not the interaction---drastically affects the behavior of matter under extreme conditions, and can determine whether the phase transition at low $T$ is a crossover or first-order. 

\section{Framework}

A practical continuum approach to the strong interaction—providing a viable alternative to numerical simulations of lattice-regularized QCD—is offered by continuum Schwinger function methods (CSMs). These methods have been successfully applied to studies of QCD at finite temperature and chemical potential, offering direct access to the infrared behavior of correlation functions and the associated dynamical phenomena; see, e.g., Refs.~\cite{roberts2000dyson, Pawlowski:2005xe, fischer2006infrared, Chen:2024emt}. The present analysis is formulated within this framework. In preparation for the results that follow, we therefore begin by outlining the Dyson–Schwinger equations (DSEs) that form the basis of our calculation.

\subsection{Quark DSE and order parameters}
\label{SEC:DSE}
At nonzero temperature and chemical potential $(T, \mu)$, the quark gap equation, or DSE for the dressed-quark propagator, can be written as (in Euclidean space)~\cite{roberts1994dyson,fischer2006infrared,roberts2000dyson,Fischer:2018sdj}
\begin{eqnarray}
    S^{-1}(\vec{p},\widetilde{\omega}_n)=i\vec{\gamma}\cdot \vec{p}+i\gamma_4\widetilde{\omega}_n+m +\Sigma(\vec{p},\widetilde{\omega_n}), 
\end{eqnarray}
with self energy given by
\begin{align}
    \Sigma(\vec{p},\widetilde{\omega_n})=&\int_{\ l,q}g^2D_{\mu\nu}(\vec{p}-\vec{q},\Omega_{nl};T,\mu)\nonumber\\
    &\times \gamma_\mu\frac{\lambda^a}{2}S(\vec{q},\widetilde{\omega}_l)\Gamma_\nu^a(\vec{p},\widetilde{\omega}_n,\vec{q},\widetilde{\omega}_l;T,\mu), 
\end{align}
where $\widetilde{\omega}_n=\omega_n+i\mu$, $\omega_n=(2n+1)\pi T$, with $n\in\mathbb{Z}$ the fermion Matsubara frequencies, $m$ is the current quark mass, $\int_{\ l,q}:=T\sum_{l=-\infty}^{\infty}\int_{-\infty}^\infty \frac{d^3q}{(2\pi)^3}$, $\Omega_{nl}=\widetilde{\omega}_n-\widetilde{\omega}_l=2(n-l)\pi T$, and the dressed-quark propagator has the general form
\begin{eqnarray}
        S^{-1}(\vec{p},\widetilde{\omega}_n)=i\vec{\gamma}\cdot \vec{p}A(\vec{p},\widetilde{\omega}_n)+i\gamma_4\widetilde{\omega}_nC(\vec{p},\widetilde{\omega}_n)+B(\vec{p},\widetilde{\omega}_n). 
\end{eqnarray}
The scalar functions, $F=A,B,C$, are complex and satisfy
\begin{eqnarray}\label{F}
F(\vec{p},\widetilde{\omega}_n)^*=F(\vec{p},\widetilde{\omega}_{-n-1}). 
\end{eqnarray}
For the gluon propagator, a particularly simple and useful model is provided by the symmetry-preserving treatment of a contact (momentum-independent) vector $\otimes $ vector quark-quark
interaction (SCI)~\cite{gutierrez2010pion},
\begin{eqnarray}
    g^2D_{\mu\nu}=\delta_{\mu\nu}\frac{4\pi\alpha_{ir}}{m_G^2}, 
\end{eqnarray}
where $\alpha_{ir}$ is the zero-momentum
value of the running-coupling constant~\cite{Cui:2019dwv,dEnterria:2022hzv,Deur:2022msf,Deur:2023dzc,Binosi:2022djx}, and $m_G$ is a gluon mass-scale that generated dynamically in QCD~\cite{Binosi:2016nme,Ding:2022ows,Ferreira:2023fva}.

We can see that, akin with the Nambu-Jona-Lasinio model~\cite{klevansky1992nambu,Oertel:2016bki}, this interaction does not have momentum dependence, so that related equations are largely algebraic, and derivations and formulae become highly transparent, which makes the outcomes easy to interpret and even enables meaningful comparisons with results from more sophisticated interactions. The physical relevance of SCI results can also be appreciated under careful interpretation. Therefore, it is widely used in many topics; for example, Refs.~\cite{Ahmad:2016iez,Cui:2017ilj,Zhang:2020ecj,Yin:2021uom,Cheng:2022jxe,Zamora:2023fgl,Yu:2024qsd,Paredes-Torres:2024mnz,Cheng:2024gyv}.

With these treatments, and taking the implicitly associated ``Rainbow approximation'', $\Gamma_\nu^a(\vec{p},\widetilde{\omega}_n,\vec{q},\widetilde{\omega}_l;T,\mu)=\frac{\lambda^a}{2}\gamma_\nu$, for the dressed quark-gluon vertex, the scalar functions satisfy,
\begin{subequations}     
\begin{align}
A(&\vec{p},\widetilde{\omega}_n) =1,\\   
C(&\vec{p},\widetilde{\omega}_n)=1 \nonumber \\
& +\frac{32\pi}{3\widetilde{\omega}_n}\frac{\alpha_{ir}}{m_G^2}\int_{\ l,q}\frac{\widetilde{\omega}_lC(\vec{q},\widetilde{\omega}_l)}{\vec{q}^2+\widetilde{\omega}_l^2C^2(\vec{q},\widetilde{\omega}_l)+B^2(\vec{q},\widetilde{\omega}_l)}, \\
B(&\vec{p},\widetilde{\omega}_n) =m \nonumber\\
&+\frac{64\pi}{3}\frac{\alpha_{ir}}{m_G^2}\int_{\ l,q}\frac{B(\vec{q},\widetilde{\omega}_l)}{\vec{q}^2+\widetilde{\omega}_l^2C^2(\vec{q},\widetilde{\omega}_l)+B^2(\vec{q},\widetilde{\omega}_l)}. 
\end{align}
\end{subequations}
Since the integrand here is $\vec{p}$-momentum independent, while the sum is $\widetilde{\omega}_n$ independent, a solution at one value of $\vec{p}$ and $\widetilde{\omega}_n$ is the solution at all values; viz., any nonzero solution must be of the form,
\begin{subequations}
\begin{align}
        &B(\vec{p},\widetilde{\omega}_n) =: R_B,\\
        &\widetilde{\omega}_n(C(\vec{p},\widetilde{\omega}_n)-1) =:  -i R_C, 
\end{align}
\end{subequations}
and Eq.~(\ref{F}) entails that $R_B$ and $R_C$ are real numbers. Then,
\begin{subequations}\label{gap1}
\begin{align}
        &R_C=\frac{8}{3\pi}\frac{\alpha_{ir}}{m_G^2}\int_{\ n,s} s^{\frac{1}{2}}\frac{ i\widetilde{\omega}_l'}{s+\widetilde{\omega}_n'^2+ R_B^2},\\
        &R_B=m+\frac{16}{3\pi}\frac{\alpha_{ir}}{m_G^2}R_B\int_{\ n,s} s^{\frac{1}{2}}\frac{ 1}{s+\widetilde{\omega}_n'^2+ R_B^2}, 
        \end{align}
\end{subequations}
where $\mu'=\mu-R_C$, $\widetilde{\omega}_n'=\omega_n+i\mu'$; $s=\vec{q}^2$ and $\int_{\ n,s}:=T\sum_{n=-\infty}^{\infty}\int_0^\infty ds$. We can see that mathematically $R_C$ is always linked with $\mu$, and plays a role in reducing it. 
Sometimes $\mu'$ is also referred to as the renormalized chemical potential~\cite{lu2015critical}.

Eq.~(\ref{gap1}) could also be written in an alternative form. Defining $E_{s,R_B}=\sqrt{s+R_B^2}$, one obtains
\begin{subequations}\label{gap2}
\begin{align}
    R_C&=\frac{4}{3\pi}\frac{\alpha_{ir}}{m_G^2}\int_{s} s^{\frac{1}{2}}[n_+-n_-],\\
    R_B&=m+\frac{8}{3\pi}\frac{\alpha_{ir}}{m_G^2}R_B\int_{s} \frac{s^{\frac{1}{2}}}{E_{s,R_B}}[n_++n_-],
\end{align}
\end{subequations}
where
\begin{align}\label{n+-}
        n_\pm&=T\sum_{n=-\infty}^{\infty}\frac{1}{E_{s,R_B}\mp i\widetilde{\omega}_n'}\nonumber\\
        &=T\sum_{n=-\infty}^{\infty}\frac{E_{s,R_B}\pm\mu'}{(E_{s,R_B}\pm\mu')^2+\omega_n^2}\nonumber\\
        &=\frac{1}{2}-\frac{1}{1+e^{\frac{1}{T}(E_{s,R_B}\pm\mu')}}.
\end{align}

To provide a quantitative way to distinguish different phases, order parameters are usually defined in phase transition studies. The quark condensate (scalar density) is one order parameter for the chiral symmetry restoration transition, defined via the dressed-quark propagator~\cite{Brodsky:2012ku,Cui:2018bor,Gao:2015kea},
\begin{align}\label{qbarq}
        \langle\bar{\psi}\psi\rangle
        &=- N_C\int_{\ n,q}tr_DS(\vec{q},\widetilde{\omega}_n)\nonumber\\
        &=- N_C\frac{3}{16\pi}\frac{m_G^2}{\alpha_{ir}}( R_B-m), 
\end{align}
where $N_C=3$ is the quark color number. On the other hand, the total quark number density (vector density) is another popular order parameter~\cite{klevansky1992nambu,Farias:2006cs,Chen:2008zr}, especially for nonzero $\mu$ areas,
\begin{align}\label{qnumber}
        \langle \psi^\dagger \psi\rangle=\langle \bar{\psi}\gamma_4\psi\rangle
        &=- N_C\int_{\ n,q}tr_D\gamma_4S(\vec{q},\widetilde{\omega}_n)\nonumber\\
        &=N_C\frac{3}{8\pi}\frac{m_G^2}{\alpha_{ir}} R_C.
\end{align}
From Eq.~(\ref{qbarq}) and Eq.~(\ref{qnumber}), we can clearly see the physical meaning of $R_B$ and $R_C$. $R_B$ is usually called the effective quark mass, while $R_C$ is closely related to quark number; hence, it is not surprising that $R_C$ has the opposite effect to that of $\mu$. 

To highlight some of the similarities and differences with the formulation of the NJL model, we present the two sets of equations side by side for comparison.

Following the formulation of the NJL model in Ref.\,\cite{lu2015critical}, the corresponding equations before regularization are
\begin{subequations}\label{NJL}
\begin{align}
    &\mu'=\mu-g_{NJL}\frac{N_f}{2\pi^2}\int_s s^\frac{1}{2}[n_+-n_-] \nonumber\\
    & =: \mu - g_{NJL}^\prime \int_s s^\frac{1}{2}[n_+-n_-]\,, \\
    &M=m+G_{NJL}\frac{N_CN_f}{\pi^2}M\int_s\frac{s^\frac{1}{2}}{E_{s,M}}[n_++n_-] \nonumber\\
    & =: m + G^\prime_{NJL} M\int_s\frac{s^\frac{1}{2}}{E_{s,M}}[n_++n_-]\,,
    \end{align}
\end{subequations}
where $M$ is the constituent mass, 
$g_{NJL}$ is the coupling constant of the four-fermion interaction, and 
\begin{eqnarray}\label{njlg}
        G_{NJL}=\frac{4N_C+1}{4N_C}g_{NJL}
\end{eqnarray}
is a renormalized coupling constant introduced in Ref.\,\cite{lu2015critical}.  One may now arrive at the following identity:
\begin{equation}
\label{njlg2}
    G^\prime_{NJL} = \tfrac{1}{2} (4 N_c+1) g^\prime_{NJL}\,.
\end{equation}

Comparing Eq.~(\ref{gap2}) and Eq.~(\ref{NJL}), one sees that these expressions exhibit a strong formal resemblance to each other. 
However, they exhibit different $N_c$ scaling.  
Indeed, defining the integral prefactors in Eq.\,\eqref{gap2} to be $g_{SCI}$, $G_{SCI}$, respectively, one has 
\begin{equation}\label{cig}
    G_{SCI} = 2 g_{SCI}\,.
\end{equation}
which should be contrasted with Eq.\,\eqref{njlg2}.
This comparison reveals a qualitative difference between a standard NJL model treatment and the SCI insofar as couplings in the $\mu$ and $M$ equations are concerned.
SCI and NJL couplings evidently carry different color factors. 
This difference originates from the fact that, in the NJL framework, the effective four-fermion coupling is taken to be color singlet from the outset, so that octet channels are not explicitly retained when performing the Fierz rearrangement, whereas in the SCI formulation the underlying color-octet interaction is preserved and treated dynamically.
At the physical value $N_c=3$, however, couplings and cutoffs can be tuned so that, within the mean-field approximation for homogeneous matter, the two approaches are practically equivalent. 

Additional significant differences emerge when one moves beyond the quark gap equation and examines, for instance, the Bethe–Salpeter equation for pseudoscalar mesons together with the associated elastic form factors. These quantities expose further discrepancies between the SCI and NJL constructions;  see, e.g., Ref.~\cite{gutierrez2010pion} and citations thereof.

\subsection{Regularization schemes}
Analogously to the NJL model, the SCI is also nonrenormalizable; therefore, a regularization scheme is unavoidable in working toward results with a meaningful interpretation. Ideally, this regularization should maximally preserve the properties of the model and symmetries~\citep{klevansky1992nambu,Farias:2006cs,gutierrez2010pion}. Currently, several schemes are available, each in some sense determines the model. For example, two of the most popular regularization schemes are the noncovariant but physically intuitive three-momentum (3M) cutoff regularization~\cite{klevansky1992nambu,Xin:2014dia,lu2015critical,inagaki2015regularization,kohyama2016parameter,Cui:2018bor,Cui:2022vsr} and a covariant regularization in proper time (PT) form~\cite{klevansky1992nambu,gutierrez2010pion,wang2013baryon,Cui:2014hya,Klahn:2015mfa,Shi:2015ufa,Wang:2016fzr,Cui:2017ilj,Zhang:2024dhs,Gifari:2024ssz,Ahmad:2016iez,Zhang:2020ecj,Yin:2021uom,Cheng:2022jxe,Zamora:2023fgl,Yu:2024qsd,Paredes-Torres:2024mnz,Cheng:2024gyv}. Hereafter, we will compare them carefully.

\subsubsection{Three-momentum cutoff regularization}
This regularization does not take into account the contributions from momenta larger than a cutoff $\Lambda_{3M}$, i.e., integrates only within $s<\Lambda_{3M}^2$. Namely, the gap equation in Eq.~(\ref{gap1}) becomes
\begin{eqnarray}\label{3M cutoff}
        R_B=m+\frac{16}{3\pi}\frac{\alpha_{ir}}{m_G^2}R_B\int_{\ n,s}^{\Lambda_{3M}^2} s^{\frac{1}{2}}\frac{ 1}{s+\widetilde{\omega}_n'^2+ R_B^2}.
\end{eqnarray}
\subsubsection{Proper time regularization}
The key treatment of this regularization is
\begin{align}\label{proper time positive}
        \frac{1}{\mathcal{T}}
        &=\int_{0}^{+\infty}dx  e^{-x \mathcal{T}} \rightarrow\int_{\tau_{uv}^2}^{\tau_{ir}^2}dx  e^{-x \mathcal{T}}\nonumber\\
        &=\frac{1}{\mathcal{T}}\times e^{-x\mathcal{T}}|^{\tau_{uv}^2}_{x=\tau_{ir}^2},if \ \mathcal{T}>0,
\end{align}
where $e^{-x\mathcal{T}}|^{\tau_{uv}^2}_{x=\tau_{ir}^2}=e^{-\tau_{uv}^2\mathcal{T}}-e^{-\tau_{ir}^2\mathcal{T}}$ and $\tau_{uv}=\frac{1}{\Lambda_{uv}},\tau_{ir}=\frac{1}{\Lambda_{ir}}$. Here, the ultraviolet cutoff $\Lambda_{uv}$ plays a role like $\Lambda_{3M}$, but is much softer, while the infrared cutoff $\Lambda_{ir}$ is introduced to mimic confinement.

For the nonzero $\mu$ case, $\mathcal{T}=s+\widetilde{\omega}_n'^2+ R_B^2=s+\omega_n^2+ R_B^2-\mu'^2+2i\omega_n\mu'$ is a complex function and requires analytic continuation to the complex plane, so we rewrite Eq.~(\ref{proper time positive}) as
\begin{align} \label{PT}
\frac{1}{\mathcal{T}} &=
\int_{0}^{+\infty}dx e^{\mp x \mathcal{T}} \rightarrow\int_{\tau_{uv}^2}^{\tau_{ir}^2} dx e^{\mp x \mathcal{T}}= \frac{1}{\mathcal{T}} \, e^{\mp x \mathcal{T}} |_{x=\tau_{uv}^2}^{\tau_{ir}^2}   \nonumber\\
&= \frac{1}{\mathcal{T}} \, e^{-\text{Sign}(\Re_\mathcal{T}) x \mathcal{T}} |_{x=\tau_{uv}^2}^{\tau_{ir}^2}=: P(\mathcal{T}).
\end{align}
where $\Re_\mathcal{T}$ is the real part of $\mathcal{T}$ and the symbol $\pm$ here corresponds to the sign of $\Re_\mathcal{T}$. There also exist some other complex forms of the PT regularization~\citep{inagaki2015regularization}, in general, they have this form,
\begin{align}\label{B}
        \frac{1}{\mathcal{T}}
        &=\frac{\mathcal{T}'}{\mathcal{T}}\times\frac{1}{\mathcal{T}'}=\frac{\mathcal{T}'}{\mathcal{T}}\int_{0}^{+\infty}dx  e^{-x \mathcal{T}'} \nonumber\\
        &\rightarrow\frac{\mathcal{T}'}{\mathcal{T}}\int_{\tau_{uv}^2}^{\tau_{ir}^2}d\tau  e^{-\tau \mathcal{T}'}
        =\frac{1}{\mathcal{T}} e^{-x\mathcal{T}'}|^{\tau_{uv}^2}_{x=\tau_{ir}^2},
\end{align}
where $\mathcal{T}'$ only needs to satisfy $\Re_{\mathcal{T}'}>0$.
One may also choose to rewrite Eq.~(\ref{B}) in many other reasonable forms, such as setting $\mathcal{T}'=|\mathcal{T}|$ or $\mathcal{T}'=i\mathcal{T}$ and so on. For example, we set  $\mathcal{T}'=Sign(\Re_\mathcal{T})\mathcal{T}$ in Eq.~(\ref{PT}). However, we have checked that different forms mainly lead to some numerical differences in the results, instead of obvious qualitative changes. Considering the intrinsic discrepancy of this model with QCD, we choose to use Eq.~(\ref{PT}) as an example for the following discussions.

\section{In Medium Analyses}
\subsection{Zero chemical potential and nonzero temperature case}
We begin by considering the zero $\mu$ and nonzero $T$ case, which can help us understand the evolution of the early universe. In this case, many equations can be simplified; for example, $\widetilde{\omega}_n=\omega_n$, $n_+=n_-$, hence $R_C=0$ GeV.

Turn to $R_B$, using the 3M regularization Eq.~(\ref{3M cutoff}), Eq.~(\ref{gap2}) becomes
\begin{align}
    & R_B =m \nonumber \\
    & \quad +\frac{8}{3\pi}\frac{\alpha_{ir}}{m_G^2}R_B\int_{0}^{\Lambda_{3M}^2} \frac{s}{E_{s,R_B}}\tanh(\frac{E_{s,R_B}}{2T})ds,
\end{align}
while using the PT regularization Eq.~(\ref{PT}), Eq.~(\ref{gap1}) becomes~\cite{wang2013baryon,Cui:2014hya}
\begin{align} 
    & R_B = m \nonumber \\
    & \quad +\frac{8}{3\sqrt{\pi}}\frac{\alpha_{ir}T}{m_G^2}R_B\int_{\tau_{uv}^2}^{\tau_{ir}^2} \frac{e^{-\tau R_B^2}}{\tau^{\frac{3}{2}}}\theta_2(e^{-4\pi^2T^2\tau}) d\tau,
\end{align}
where $\theta_2(q)=2q^\frac{1}{4}\sum_{n=0}^{+\infty}q^{n(n+1)}$ is the Jacobi theta-function.

\begin{figure}[t]
    \includegraphics[width=\columnwidth]{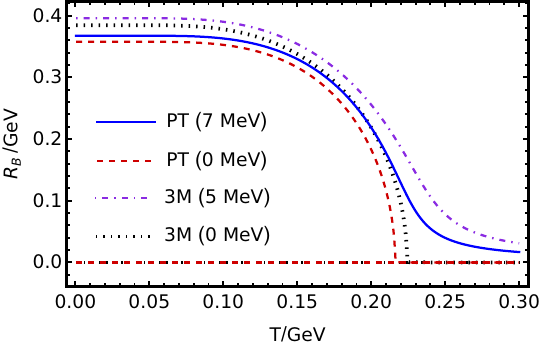}
\caption{Solutions of the gap equation at $\mu=0$ GeV using PT (solid blue/dashed red curve for $m=0.007/0$ GeV) and 3M (dot-dashed purple/dotted black curve for $m=0.005/0$ GeV) regularizations.\label{result1}}
\end{figure} 

\begin{table}[t]
\caption{\label{parameters}%
Regularization and interaction strength parameters used herein, whose values were chosen as described in the text. All quantities listed are measured in GeV, except for the coupling constant $\alpha_{ir}$, which is dimensionless.
}
\begin{ruledtabular}
\begin{tabular}{cccccc}
\textrm{Schemes}&
\textrm{$m$}&
\textrm{$\alpha_{ir}$}&
\textrm{$m_G$}&
\textrm{$\Lambda_{3M}/\Lambda_{uv}$}&
\textrm{$\Lambda_{ir}$}\\
\colrule
    3M	& 0.005 & $0.802\pi$ & 0.8 &     0.650   &  -   \\
    PT	& 0.007 & $0.93\pi$  & 0.8 &    0.905   & 0.24 \\
\end{tabular}
\end{ruledtabular}
\end{table}
\subsection{Zero temperature and nonzero chemical potential case}
Compact stars with their small ratio of temperature to chemical potential, $T/\mu\approx 0$, are well approximated by $T\rightarrow0$ GeV and nonzero $\mu$. In this case, the Matsubara frequencies become continuous, i.e., $\omega_n\rightarrow p_0$, and the sum then becomes an integral over the real axis,
\begin{eqnarray}
        \lim_{T\rightarrow 0}T\sum_{n=-\infty}^{\infty}f(\omega_n)=\frac{1}{2\pi}\int_{-\infty}^{\infty}dp_0 f(p_0),
\end{eqnarray}

so

\begin{eqnarray}
        R_C&=&\frac{4}{3\pi}\frac{\alpha_{ir}}{m_G^2}\int_{0}^{\max\{0,\mu'^2-R_B^2\}}s^{\frac{1}{2}}ds \nonumber \\
        &=&\left\{
        \begin{array}{ll}
            0, &\mu'<R_B,\\
            \frac{8}{9\pi}\frac{\alpha_{ir}}{m_G^2}(\mu'^2-R_B^2)^{\frac{3}{2}}, &\mu'>R_B,
        \end{array}
        \right. 
\end{eqnarray}

which gives the total quark number density as
\begin{eqnarray}
        \langle q^\dagger q\rangle
        &=&
        \left\{
        \begin{array}{ll}
            0, &\mu'<R_B,\\
            N_C\frac{(\mu'^2-R_B^2)^{\frac{3}{2}}}{3\pi^2}. &\mu'>R_B.
        \end{array}
        \right.
\end{eqnarray}
consistent with Ref.~\citep{PhysRevD.58.096007}.

Turn to $R_B$, using Eq.~(\ref{n+-}) and the 3M regularization Eq.~(\ref{3M cutoff}), Eq.~(\ref{gap2}) becomes
\begin{eqnarray}\label{3mmu}
    R_B=m+\frac{8}{3\pi}\frac{\alpha_{ir}}{m_G^2}R_B\int_{\max\{0,\mu'^2-R_B^2\}}^{\Lambda_{3M}^2}\frac{s^{\frac{1}{2}}}{\sqrt{s+R_B^2}}ds,
\end{eqnarray}
while using the PT regularization Eq.~(\ref{PT}), Eq.~(\ref{gap1}) becomes
\begin{eqnarray}\label{ptmu} 
    R_B=m
    +\frac{8}{3\pi^2}\frac{\alpha_{ir}}{m_G^2}R_B\int_0^\infty ds\int_{-\infty}^\infty dp_0 s^{\frac{1}{2}}P(\mathcal{T}),
\end{eqnarray} 
where $\mathcal{T}=s+p_0'^2+ R_B^2$ and $p_0'=p_0+i\mu'$, hence $\Re_\mathcal{T}=s+R_B^2+p_0^2-\mu'^2$.

\section{Results}

We solve the gap equations using the parameters in Table~\ref{parameters}. These values are chosen in order to draw close comparison with NJL results. We set $\alpha_{ir}=0.802 \pi$ for the 3M regularization used in Ref.~\cite{lu2015critical} with $N_C=3, N_f=2$ so that $G_{SCI}=G_{NJL}\approx 5.5\ \mathrm{GeV^{-2}}$ and then slightly retune
$\Lambda_{3M}$ from 0.631 GeV to 0.65 GeV to compensate for the coefficient mismatch between Eq.~(\ref{njlg2}) and Eq.~(\ref{cig}), thereby obtaining results analogous to those of the NJL results.
PT parameters are fixed analogously to reproduce the correct vacuum solution,
ensuring that in-medium differences reflect the impact of different regularization schemes rather than parameter tuning in vacuum. 

\begin{figure}[t]
\centering
\begin{overpic}[width=\columnwidth]{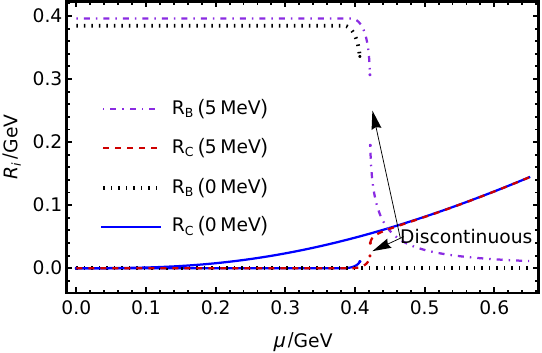}
    \put(2,65){\textsf{\large A}} 
\end{overpic}
\\
\begin{overpic}[width=\columnwidth]{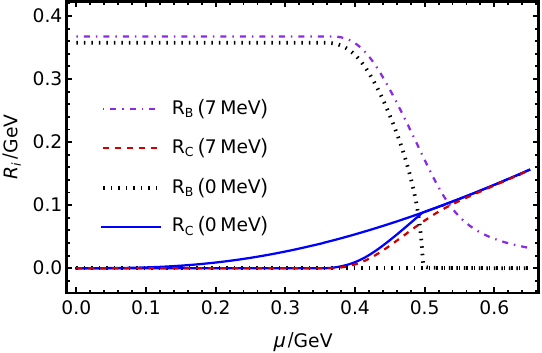}
    \put(2,65){\textsf{\large B}} 
\end{overpic}
\caption{Solutions of the gap equation with different regularization schemes at $T\rightarrow0$ GeV. Panel A. 3M regularization; Panel B. PT regularization. Both: the dot-dashed purple and dashed red curves show $R_B$ and $R_C$ beyond the chiral limit, respectively, while the dotted black and solid blue curves represent the corresponding results in the chiral limit $m=0$ GeV.\label{result2}}
\end{figure}

Solutions of the gap equation with different regularization schemes are shown in Fig.~\ref{result1} and Fig.~\ref{result2}, respectively.

We can see clearly from Fig.~\ref{result1} that, in the chiral limit $m=0$ GeV case, the gap equation always has a 
solution with restored chiral symmetry, 
i.e. $R_B=0$ GeV, which is usually named the Wigner solution; see, e.g.,~\cite{Wang:2016fzr,Cui:2018bor,Qin:2010nq}. Another solution is the chiral symmetry broken Nambu solution, i.e. $R_B\neq 0$ GeV. 
In domains where the Nambu and Wigner solutions coexist, the thermodynamically stable branch is the one that minimizes the thermodynamic potential. At low temperature this is the Nambu solution.
Increasing the temperature, the system undergoes a second-order phase transition to the Wigner phase.~\cite{Cui:2018bor,wang2013baryon,roberts2000dyson} However, once we go beyond the chiral limit, there is only the Nambu solution, and the transition becomes a crossover, which has been discussed widely in the literature~\cite{wang2013baryon,Aoki:2006we,Bazavov:2011nk,Fukushima:2010bq,Bhattacharya:2014ara}. Moreover, it is interesting to note that 3M and PT regularization schemes give very similar results, both qualitatively and quantitatively.

Next, we consider the $T\rightarrow0$ GeV case, as shown in Fig.~\ref{result2}. 
In the 3M regularization, as shown in Fig.~\ref{result2}A, we can clearly see that there is a first-order phase transition. When going beyond the chiral limit, the Nambu and Wigner solutions coexist in the region from $\mu\approx 421.5$ MeV to $\mu\approx 422.0$ MeV. 
For the PT regularization scheme, the picture changes significantly. As shown in Fig.~\ref{result2}B, the pattern is almost the same as that in Fig.~\ref{result1}, besides $R_C$ becoming nonzero and increasing gradually with $\mu$. Namely, in this scheme, the $\mu$ effect is very similar to the $T$ one.
 From Fig.~\ref{result2}, we also notice that $R_C=0$ GeV is a good approximation at low $\mu$; while when $\mu$ is large, $R_C$ effect is not negligible.

Comparing Fig.~\ref{result1} and Fig.~\ref{result2}, we find that the qualitative effect of increasing $T$ at zero $\mu$ is similar for both schemes, whereas with increasing $\mu$ at low $T$ the regularization scheme determines the order of the phase transition.

\begin{figure}[t]
\centering
\includegraphics[width=\columnwidth]{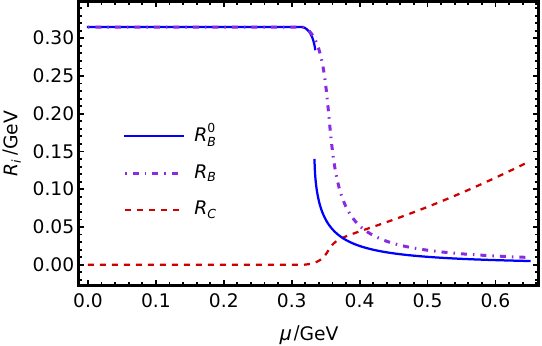}
\caption{$R_C$ effect on the phase transition in the 3M regularization scheme. Legend: solid blue curve - ignore the effects of $R_C$ using parameters from Table~\ref{parameters} and set $\alpha_{ir}=0.73\pi$; dotdashed purple curve - the result of $R_B$ restoring the effects of $R_C$; dashed red curve - the result of $R_C$.\label{c0}}
\end{figure}  

\section{Discussion}
To clearly show the influence of the vector mean field $R_C$, we use the 3M parameters from Table~\ref{parameters} and set $\alpha_{ir}=0.73\pi$. As shown in Fig.~\ref{c0}, this turns the previously observed first order phase transition from Fig.~\ref{result2}A into a crossover. If we set $R_C=0$ GeV, the curve will move to lower $\mu$ and turn first-order again. Obviously, this shows that $R_C$ has a significant impact on the order of the phase transition and weakens the first order phase
transition.

The key point originates from the minus sign in the effective chemical potential $\mu'=\mu-R_C$, which reflects the repulsive nature of the vector interaction among quarks~\cite{Buballa:2003qv,klevansky1992nambu}. Since $R_C\propto \langle \psi^\dagger \psi\rangle$ grows with the quark number density, the repulsion increases as the system becomes denser. Consequently, achieving a further increase in density requires a larger external chemical potential to compensate for this enhanced internal repulsion. In other words, $R_C$ systematically suppresses the influence of the chemical potential on the thermodynamic potential. This mechanism naturally leads to a weakening of the first-order chiral phase transition at high density.

\begin{figure}[tb]
\begin{overpic}[width=\columnwidth]{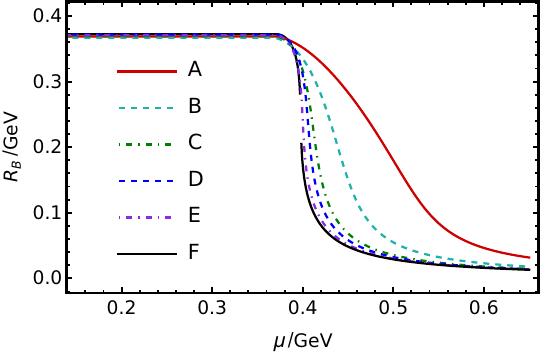}
    \put(2,65){\textsf{\large A}} 
\end{overpic}

\begin{overpic}[width=\columnwidth]{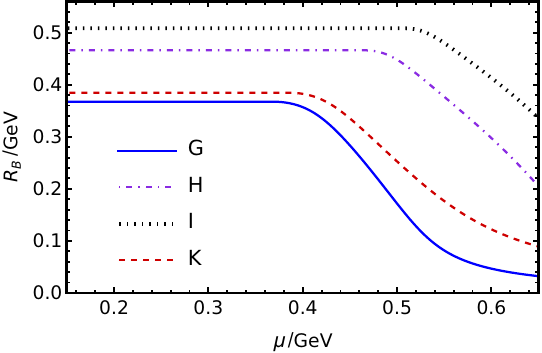}
    \put(2,65){\textsf{\large B}} 
\end{overpic}
\caption{Comparison of solutions for different parameter sets. Panel A. Using 6 different parameters listed in the upper panel of Table~\ref{group sets}. Panel B. Using 4 different parameters listed in the lower panel of Table~\ref{group sets}.\label{fig:comparison}}
\end{figure}

\begin{table}[t]
\caption{Parameter sets used in Fig.~\ref{fig:comparison}. Upper panel - 6 sets of $\Lambda_{uv}$ and $\Lambda_{H}$ for Eq.~(\ref{rbli}). Lower panel - 4 parameter sets for PT regularization. All quantities listed except for the dimensionless coupling constant $\alpha_{ir}$ are measured in GeV.\label{group sets}}
\begin{ruledtabular}
\begin{tabular}{ccccc}
Cases   &\textrm{$\Lambda_{uv}$}  &\textrm{$\Lambda_{H}$}&\textrm{$\alpha_{ir}$}&\textrm{$m$}\\
\colrule
A   &0.905  &10     &$0.73\pi$&0.005\\
B   &1.5    &0.787  &$0.73\pi$&0.005\\
C   &3      &0.67   &$0.73\pi$&0.005\\
D   &5      &0.64   &$0.73\pi$&0.005\\
E   &10     &0.62   &$0.73\pi$&0.005\\
F   &100    &0.605  &$0.73\pi$&0.005\\
\colrule
G   &0.905  &-&$0.93\pi$&   0.007\\
H   &0.905  &-&$1.15\pi$&   0.007\\
I   &1      &-&$0.93\pi$&   0.007\\
K   &0.905  &-&$0.93\pi$&   0.02\\
\end{tabular}
\end{ruledtabular}
\end{table}

Next, we analyze how the regularization schemes change the order of the phase transition. 
As mentioned above, since both models are nonrenormalizable, one has to introduce some special treatment to cure the ultraviolet divergence and make the quantities finite. Comparing Eq.~(\ref{3mmu}) and Eq.~(\ref{ptmu}), we find that the 3M regularization scheme imposes a hard cutoff $\Lambda_{3M}^2$ 
which eliminates contributions from higher momenta. Contrarily, the momentum integration goes up to $+\infty$ in the PT regularization, with a soft cutoff due to the exponential damping factor $e^{-\tau^2\mathcal{T}}$.

To interpolate between hard cutoff and the soft exponential damping factor, we introduce the following equation,
\begin{eqnarray}\label{rbli}
        R_B=m
        +\frac{8}{3\pi^2}\frac{\alpha_{ir}}{m_G^2}R_B\int_{0}^{\Lambda_{H}^2} ds\int_{-\infty}^{\infty} dp_0 s^{\frac{1}{2}}P(\mathcal{T}),
\end{eqnarray}
where $\mathcal{T}=s+p_0'^2+ R_B^2$. 
The exponential factor related to $\tau_{ir}$ is approximately zero; we ignore it for this discussion. Using the PT parametrization from Table~\ref{parameters} we vary $\Lambda_{uv}$ as shown in Table~\ref{group sets}. 
In order to keep the vacuum mass $R_B(T=\mu=0\ \mathrm{GeV})$ constant for all $\Lambda_{uv}$ we adjust $\Lambda_{H}$ accordingly. In this way we interpolate from a PT type solution (A) to a 3M type solution (F). Fig.~\ref{fig:comparison}A shows how at $T=0$ GeV the PT crossover in $R_B(\mu)$ then gradually changes into a first order phase transition.

We conclude that, as a noncovariant regularization scheme, the 3M cutoff treatment introduces a hard and sharp cutoff in momentum space, which tends to result in more abrupt changes in the behaviors of the physical quantities (we use effective mass as an example, and the similar pattern appears in related quantities), expressed as a first-order phase transition.
While the PT regularization is a covariant scheme that smoothly regularizes the interaction by introducing an exponential damping factor, which is gentle and soft, so that most related quantities and processes will also tend to change continuously, expressed as a crossover.

\begin{figure}[t]
\includegraphics[width=\columnwidth]{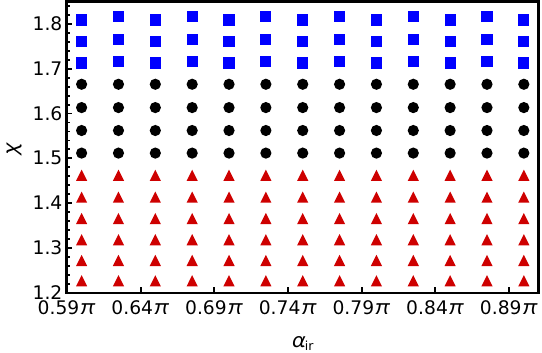}
\caption{The size of the dimensionless parameter $\chi$, Eq.\,\eqref{Xdef}, controls the order of the transition at $m=0.005$ GeV. Legend: blue squares – first-order for both $R_C=0$ GeV and $R_C\neq0$ GeV; black circles – first-order only at $R_C=0$ GeV and crossover at $R_C\neq0$ GeV; red triangles – crossover in both cases.\label{Fig7}}
\end{figure}  

Finally, we discuss the influence of the remaining parameters on the phase transition as listed in the lower panel of Table~\ref{group sets}. The results are shown in Fig.~\ref{fig:comparison}B. With PT regularization, any combination of $\Lambda_{uv}$, $\alpha_{ir}$ and $m$ results in a crossover phase transition at  $T\rightarrow0$ GeV; for the 3M regularization, we show numerically that the dimensionless quantity 
\begin{equation}
    \chi:=\frac{\alpha_{ir}\Lambda^2_{3M}}{m_G^2} \label{Xdef}
\end{equation} 
determines the order of the phase transition. Fig.~\ref{Fig7} shows that smaller values of $\chi\lesssim1.7$ lead to a crossover at $m=0.005$ GeV, whereas $\chi\gtrsim1.7$ produce a first-order transition. Although one may set $R_C=0$ GeV, its effect is merely to shift the numerical value of $\chi$ down without altering this qualitative behavior.


\section{Conclusions}
We used a contact interaction to compare the effects of different regularization schemes: the noncovariant three-momentum cutoff and the covariant proper time treatment, at nonzero temperature $T$ or chemical potential $\mu$. 
At zero $\mu$ and nonzero $T$, the chiral phase transition is a smooth crossover for both regularization schemes. 
When turning to the low $T$ and nonzero $\mu$ region, different regularization schemes and parametrizations may provide qualitatively distinct behavior, namely, crossover or first-order. 
We find that different parameter choices produce only quantitative changes in the covariant regularization, yet they determine the order of the phase transition in the noncovariant case.

This sensitivity is problematic. In a renormalizable quantum field theory, the phase structure is expected to be independent of the regularization procedure if it has been carried out consistently.
In this study, we kept vacuum observables and the underlying interaction fixed. 
In conclusion, for the non-renormalizable contact model the regulator is effectively part of the physics, and predictions concerning the phase structure in medium are intrinsically scheme dependent.

At the same time, the bare-vertex approximation violates the Ward identity at nonzero chemical potential, indicating that adopting something like Ball–Chiu vertex is a natural step for future refinement.

\begin{acknowledgments}
We thank C.\ D.\ Roberts for valuable suggestions.
Work supported by:
Natural Science Foundation of Jiangsu Province, grant no.\ BK20220122;
National Natural Science Foundation of China, grant no.\ 12233002.
\end{acknowledgments}

\bibliography{li25}

\end{document}